\documentclass[conference,letterpaper]{IEEEtran}
\usepackage{graphicx}

\usepackage{amssymb}
\usepackage{lscape}
\usepackage{longtable}

\usepackage[cmex10]{amsmath}
\usepackage{color}
\usepackage{fancyhdr}
\usepackage[caption=false,font=footnotesize]{subfig}

\renewcommand{\thispagestyle}[2]{} 
\fancypagestyle{plain}{
        \fancyhead{}
        \fancyhead[C]{first page center header}
        \fancyfoot{}
        \fancyfoot[C]{first page center footer}
}
\begin{document}

%
\title{Role of Bloom Filter in Big Data Research: A Survey}

\author{\IEEEauthorblockN{Ripon Patgiri}
\IEEEauthorblockA{Dept of Computer Science \& Engineering\\
National Institute of Technology Silchar\\
Assam-788010, India\\
Email: ripon@cse.nits.ac.in}
\and
\IEEEauthorblockN{Sabuzima Nayak}
\IEEEauthorblockA{Dept of Computer Science \& Engineering\\
National Institute of Technology Silchar\\
Assam-788010, India\\
Email: sabuzimanayak@gmail.com}
\and
\IEEEauthorblockN{Samir Kumar Borgohain}
\IEEEauthorblockA{Dept of Computer Science \& Engineering\\
National Institute of Technology Silchar\\
Assam-788010, India\\
Email: samir@nits.ac.in}
}

\maketitle

\begin{abstract}
Big Data is the most popular emerging trends that becomes a blessing for human kinds and it is the necessity of day-to-day life. For example, Facebook. Every person involves with producing data either directly or indirectly. Thus, Big Data is a high volume of data with exponential growth rate that consists of a variety of data. Big Data touches all fields, including Government sector, IT industry, Business, Economy, Engineering, Bioinformatics, and other basic sciences. Thus, Big Data forms a data silo. Most of the data are duplicates and unstructured. To deal with such kind of data silo, Bloom Filter is a precious resource to filter out the duplicate data. Also, Bloom Filter is inevitable in a Big Data storage system to optimize the memory consumption. Undoubtedly, Bloom Filter uses a tiny amount of memory space to filter a very large data size and it stores information of a large set of data. However, functionality of the Bloom Filter is limited to membership filter, but it can be adapted in various applications. Besides, the Bloom Filter is deployed in diverse field, and also used in the interdisciplinary research area. Bioinformatics, for instance. In this article, we expose the usefulness of Bloom Filter in Big Data research. 
\end{abstract}


\IEEEpeerreviewmaketitle

\section{Introduction}
Big Data has been revolutionizing the IT industry since its inception, and it is proven as a game changer paradigm. Nowadays, Big Data is continuously dominating IT industry, because data are oils of the modern era.  Interestingly, Big Data touches almost all fields, including Government sector, Business, Politics, Science, Economy, and Healthcare \cite{RA}. Therefore, datacenter deals with a variety of data from various sources. The varieties of data are typically classified  into three category, namely, structured, semi-structured, and unstructured data which compose a large sized data silo. Revenue and investments are growing, and it is expected to grow till 2030. Big Data is a data silo of variety of data that grows exponentially. From the year 2003, Big Data grows exceptional pace and continue to grow with the same pace. Moreover, Big Data gears up the speed speed of growth in 2009. In 2012, IDC predicted that Universal Data size should touch 44 ZB in 2020. In most of the IT industries, petabytes of data are a dime a dozen. As we know that data are generated through various sources, for instance, IoT. Various organizations are gathering data to generate revenue. Because, data are an asset to the organizations. Most of the companies have surpassed petabytes of data, and they have already landed in exabytes of data. For example, Google and NSA reach 15 EB and 10 EB data. Undoubtedly, it is an onerous task to deal with this kind of data sizes. Thus, Big Data is defined by $V_3^{11}+C$ \cite{RA}. 

\begin{figure*}[!ht]
\centering
\includegraphics[width=0.8\textwidth]{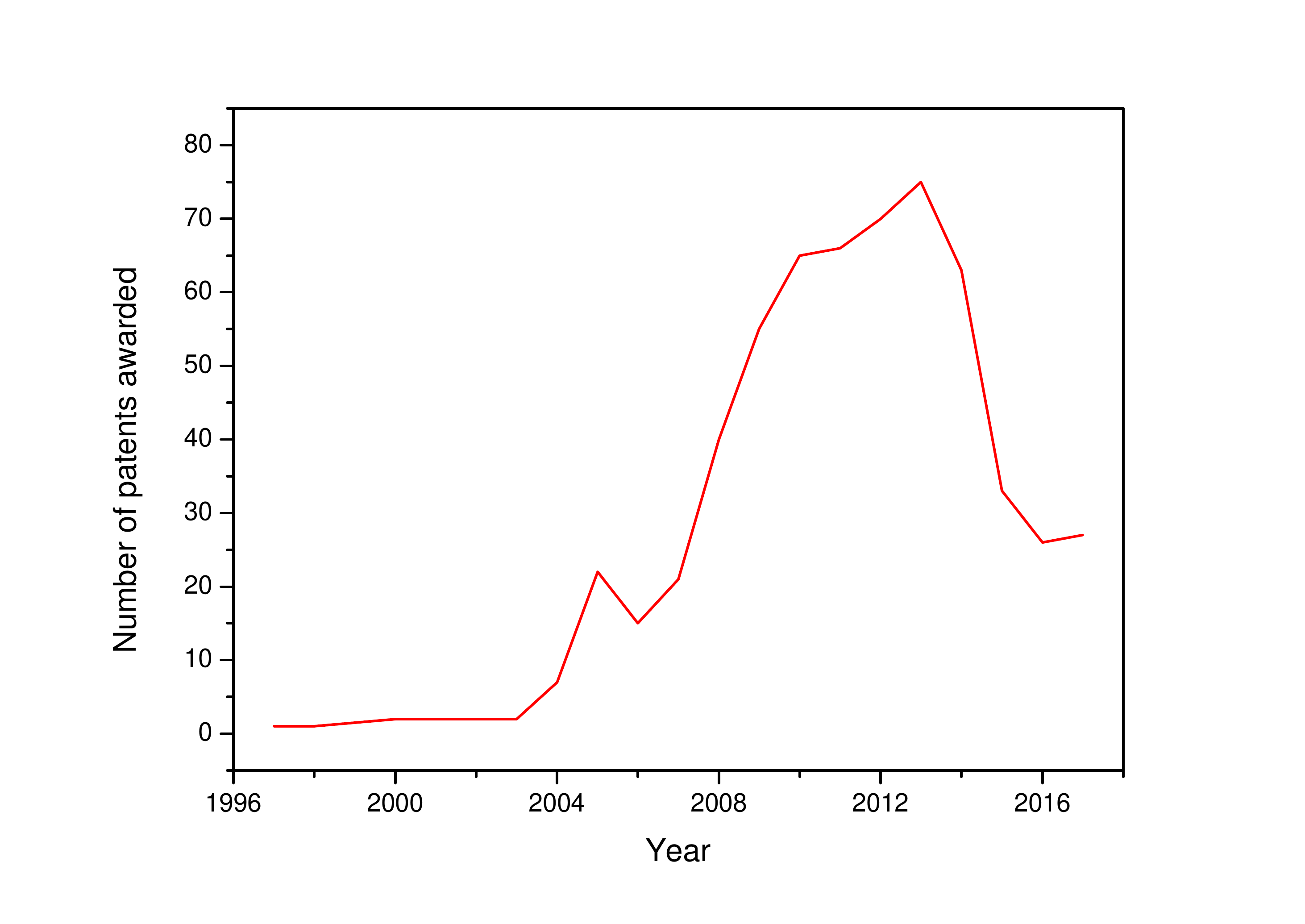}
\caption{Number of patent awarded in Bloom Filter/using Bloom Filter according to Google Patent Search Engine.}
\label{pat}
\end{figure*}

Undoubtedly, managing a large data silo is a grand challenge. There are dozens of issues and challenges in Big Data \cite{RP}. Therefore, there are numerous research papers have been published to address various issues of Big Data by various publishers, for example, Elsevier, ACM, IEEE, and Springer. Similarly, Bloom Filter \cite{Bloom} also applied in numerous filed and published numbers of papers by various publishers. Figure \ref{pat} extrapolate the number of patents awarded based on Bloom Filter according to Google Patent Search Engine. In other word, Bloom Filter is very popular hashing technique. In addition, Bloom Filter is a very simple yet powerful data structure to handle large scale data by sacrificing a small amount of memory space. Unlike a conventional hash data structure, Bloom Filter is more space efficient and Bloom Filter is able to reduce memory space consumption by order of magnitude. The Bloom Filter is an approximate membership query. Therefore, the Bloom Filter is unable stand itself, but it is just like a cog in a machine. Big Data can be enhanced in the  presence of Bloom Filter.  

In this article, we present the impact of Bloom Filter in Big Data. We also outline future challenge of Bloom Filter in Big Data. In Section ~\ref{DS}, we explore the employment of Bloom Filter in Big Data storage system. Section~\ref{Bloom} outlines Bloom Filter. Section~\ref{DS} reviews the employment of Bloom Filter in Big Data storage system. In addition, comparative study is presented in Section ~\ref{DS}. Section~\ref{FD} highlights the future direction of Bloom Filter in Big Data storage system. And finally, article is concluded by Section~\ref{Con}. 

\section{Bloom Filter}
\label{Bloom}
\begin{figure}[ht]
\centering
\includegraphics[width=0.45\textwidth]{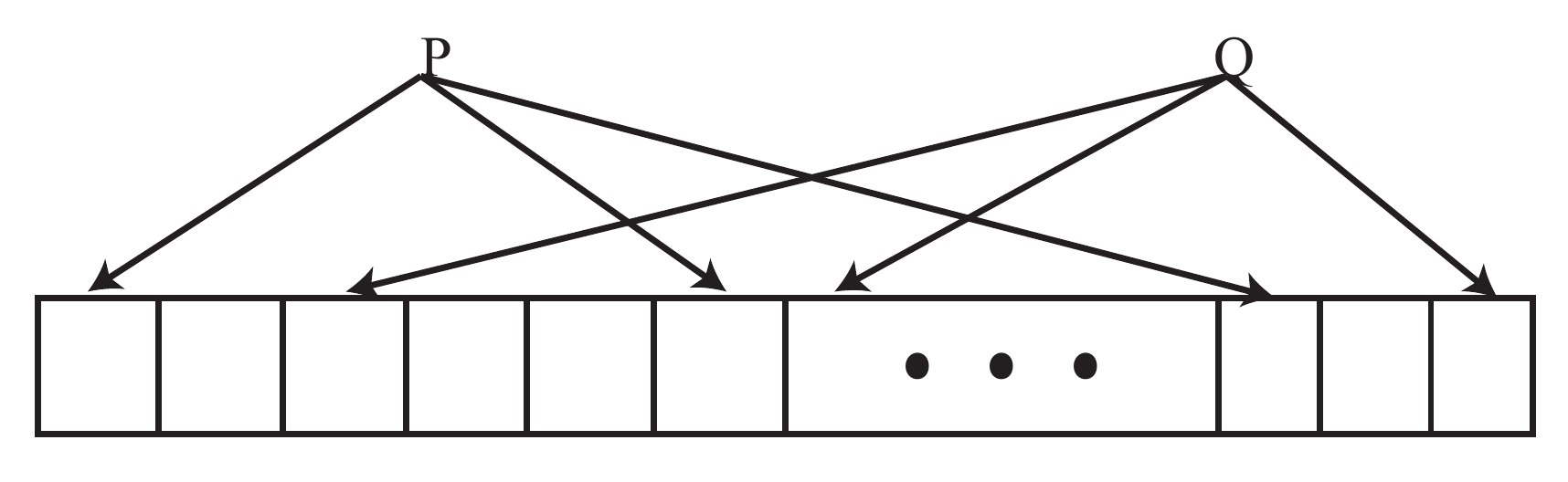}
\caption{Conventional Bloom Filter.}
\label{bf}
\end{figure}

Bloom Filter is a probabilistic data structure with some errors. Bloom Filter is a variant of hashing algorithm that uses a tiny amount of extra memory space which is introduced by Burton Howard Bloom \cite{Bloom}. Bloom Filter has been applying various research area to boost up the performance of a system. Bloom Filter is a membership filtering algorithm that returns either ``true'' or ``false''. However, the ``true'' can be either false positive or true positive. Similarly, the ``false'' can also be either false negative or true negative. The false positive and false negative are the error of Bloom Filter. Nevertheless, the error is negligible and tolerable. But, Bloom Filter does not suit for many systems too. Real-time system, for instance. 

Let us assume, $P$ and $Q$ are two elements to be inserted into the Bloom Filter. Bloom Filter depends on the $k$ value which is the number of hash functions of a key or an element. If $k=3$, then Figure~\ref{bf} depicts the number of cells occupied by $P$ and $Q$. $P$ is hashed into three different places in the array. Some variants of Bloom Filter store fingerprints of the keys, and some variants just put '1' or '0'. Similarly, $Q$ is also hashed into three different locations in the array. Those three positions must be set (true) while checking the membership of $P$ or $Q$ and it is a necessary condition for positive answer of the Bloom Filter.

Let, $S$ be a set and $K$ be the elements of $S$ where $K\in S$, $K=K_1,K_2,K_3,\ldots,K_n$, and $n$ is the total number of elements. Let, $B$ be the Bloom Filter and insert all elements of $S$ into $B$. Let us assume, $K_j$ be a random query element where $j=1,2,3,\ldots$. Article \cite{DDoS} defines false positive, true positive, false negative, and true negative as follows-
\begin{itemize}
\item False positive: If $K_j\not\in S$ and Bloom Filter returns $K_j\in B$, then the result of Bloom Filter is a false positive.
\item True positive: If $K_j\in S$ and Bloom Filter returns $K_j\in B$, then the result of Bloom Filter is a true positive.
\item False negative: If $K_j\in S$ and Bloom Filter returns $K_j\not\in B$, then the result of Bloom Filter is a false negative.
\item True negative: If $K_j\not\in S$ and Bloom Filter returns $K_j\not\in B$, then the result of Bloom Filter is a true negative.
\end{itemize}

The key issue in Bloom Filter is false positive and false negative. Most of the Bloom Filter variants suffer from false positive, but not false negative. However, the false negative issue arises in the deletion of an element. The counting Bloom Filters are introduced to address the issue of scalability. But, false negative is a serious issue with the counting Bloom Filter upon deleting an item. Therefore, most of the Bloom Filters avoid deletion of a key. On the other hand, Cuckoo Filter \cite{Fan} reduces the false positive, however, it suffers from performance issue. Also, Rottenstreich et al. \cite{Ori} implements variable counting Bloom Filter, called CBF. CBF avoids increment by one on insertion to reduce false positive. CBF uses $B_h$ sequence to increase upon insertion and decrease upon deletion \cite{Ori}. However, memory consumption per item is very high. 

Let us outline the false positive of conventional Bloom Filter. Let, $m$ be the total memory occupied by the Bloom Filter, $n$ be the total element entry into the Bloom Filter and $k$ be the total number of hash functions, then the probability of a bit to be `0' after $n$ element insertion in the given filter is \cite{Almeida} \[\left(1-\frac{1}{m}\right)^{nk}\]
The probability of the given bit to be `1' is
\[\left(1-\left(1-\frac{1}{m}\right)^{nk}\right)\approx \left(1-e^{-kn/m}\right)\]
The probability of all bits to be `1' is 
\[\left(1-\left(1-\frac{1}{m}\right)^{nk}\right)^k\approx \left(1-e^{-kn/m}\right)^k\]

On the other hand, F. Grandi \cite{Grandi} present another approach to analyze the false positive probability. Let, $X$ be the random variable representing the total number of set bits (i.e., `1'), then
\[E[X]=m\left(1-\left(1-\frac{1}{m}\right)^{nk}\right)\]
Let us conditioned the false positive to a number $X=x$, then
\[Pr(FP|X=x)=\left(\frac{x}{m}\right)^k\]
where $x$ is set bits. Then, F. Grandi \cite{Grandi} calculates the false positive probability (FPP) as follows
\[FPP=\sum_{x=0}^m Pr(FP|X=x)~Pr(X=x)\]
\[=\sum_{x=0}^m f(x)\left(\frac{x}{m}\right)^k\]
where $f(x)$ is probability mass function. F. Grandi \cite{Grandi} applies $\gamma-tranformation$ to obtain the value of probability mass function, and the false positive probability of standard Bloom Filter is calculated as follows
\[FPP=\sum_{x=0}^m \left(\frac{x}{m}\right)^k {m\choose x} \sum_{j=0}^x \left(-1\right)^j {x\choose j} \left(\frac{x-j}{m}\right)^{kn}\]

Thus, the probability of false positive also depends on $m$ and $n$. The false positive becomes overhead of Bloom Filter. If a system can tolerate negligible overhead, then Bloom Filter can enhance a performance of a system. Therefore, Bloom Filter is deployed to various domains, namely, Big Data \cite{BigTable}, Deduplication \cite{BloomFlash,Zhang,JLi}, Network Security \cite{DDoS,halagan}, Network Traffic control \cite{Mal}, Name Lookup \cite{He,Dai}, IP address lookup \cite{Lin,Mun}, Biometric \cite{Biomet,Biomet1}, Bioinformatics \cite{Heo,Abyss}, File System \cite{Wei}, Indexing, and many more. However, Bloom Filter is not suitable in case of correct query-answer requirements. For instance, password database. Because, Bloom Filters response a query by approximating either true or false.

\subsection{Classifications of Bloom Filter}
\begin{figure}
\centering
\includegraphics[width=0.45\textwidth]{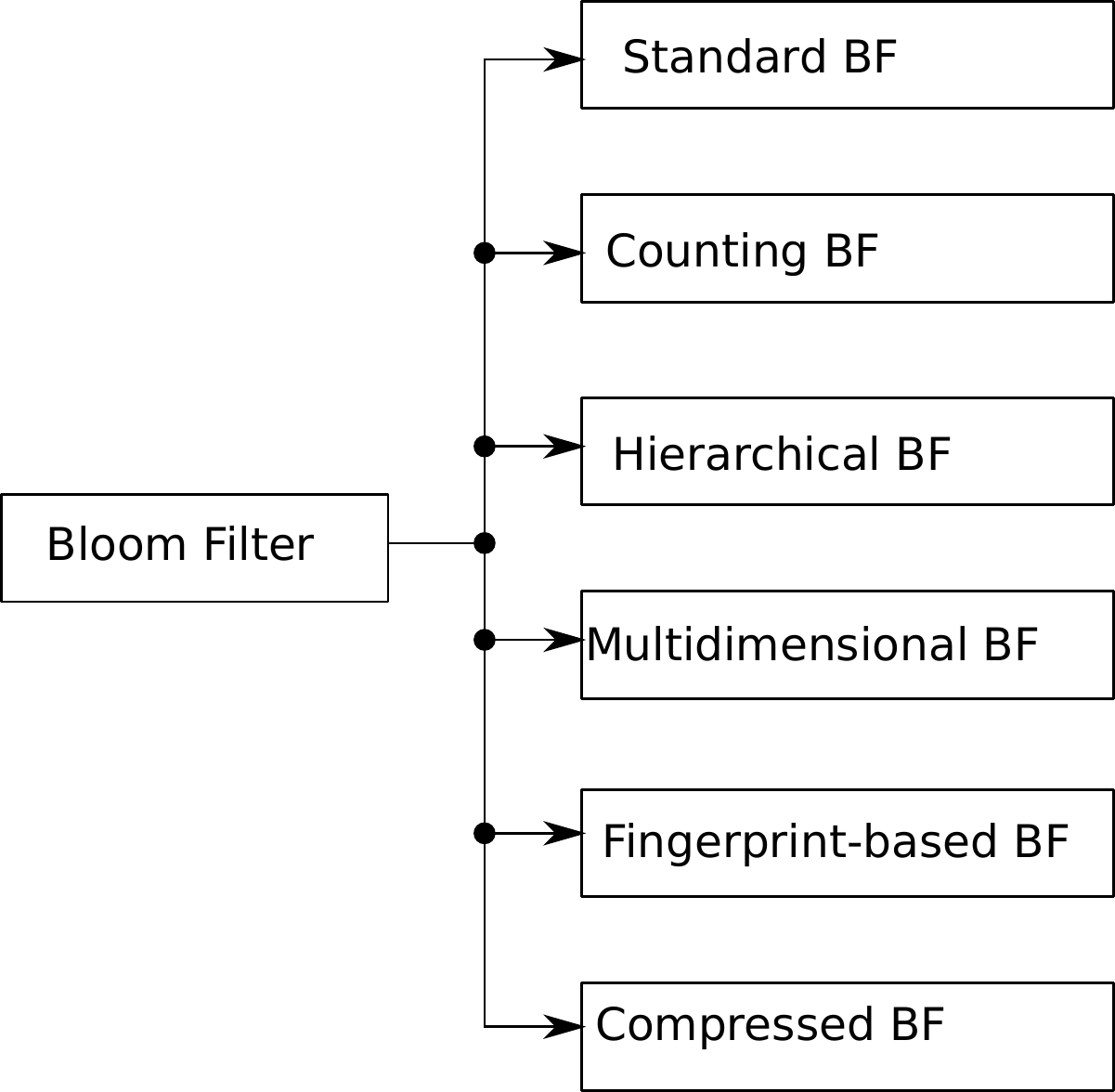}
\caption{Taxonomy of Bloom Filter}
\label{taxo}
\end{figure}

Bloom Filter is classified into six major categories as sketched in Figure ~\ref{taxo}, namely, Standard Bloom Filter, Counting Bloom Filter, Hierarchical Bloom Filter, Multidimensional Bloom Filter, Fingerprint-based Bloom Filter, and Compressed Bloom Filter.

\subsubsection{Standard Bloom Filter}
Standard Bloom Filter (SBF) is known as conventional Bloom Filter. There are false positive and false negative in SBF. False negative occurs in case of deleting some elements, since, Bloom Filter is a stateless filter. Therefore, SBF cannot guarantee the false negative. Also, false positive of SBF depends on number of hash functions which is termed as $k$. The value of $k$ cannot be too large or too small. The optimal value of $k$ is defined by $k=\frac{m}{n}~ln2$.
\subsubsection{Counting Bloom Filter}
Counting Bloom Filter (CBF) emerges due to lack of scalability in SBF. CBF is able to scale very high number of inputs. CBF comprises of many counters in the array, and the counters count the number of inputs. Thus, CBF is able to scale very large amount of input as compared to SBF. The counters are incremented in the insertion of an input item and decremented in the deletion of an input item. Therefore, there is very high false positive probability as compared to SBF. Also, CBF causes false negative due to deletion. 
\subsubsection{Fingerprint Bloom Filter}
CBF has a high false positive probability. Fingerprint-based CBF has been introduced to reduce the false positive probability. For instance, Cuckoo Filter \cite{Fan}. Fingerprint-based CBF stores a small sized hash value instead of dealing with binary bits. Thus, a fingerprint can directly be matched with a stored fingerprint which reduces false positive probability. However, the space consumption is more than CBF.
\subsubsection{Hierarchical Bloom Filter}
The hierarchical Bloom Filters (HBF) are known as tree or forest-structured Bloom Filter. The HBF is a Bloom Filter by augmenting several Bloom Filter to achieve high accuracy, scalability, and low false positive probability. For instance, Forest Structured Bloom Filter \cite{Lu} and BloomStore\cite{GLu}. HBF is deployed in deduplication of very large scale datasets. Moreover, HBF uses Flash-memory or HDD to increase scalability.
\onecolumn
\begin{landscape}
\begin{longtable}{p{3cm}p{2.5cm}p{1cm}p{1cm}p{1cm}p{1cm}p{2cm}p{2cm}p{2cm}}
\caption{Bloom Filter in a large scale membership filtering}\\
\hline
\textbf{Author or Name} &  \textbf{Name of Bloom Filter} & \textbf{False Positive} & \textbf{False negative} & \textbf{Scala-bility} & \textbf{Delete} & \textbf{Type} & \textbf{Platform} & \textbf{Purposes}
\\
\hline
BigTable\cite{BigTable} & Conventional & $\checkmark$ & $\times$ & $\times$ & $\times$ & Conventional & RAM \& HDD & Database\\

BloomFlash~\cite{BloomFlash}  & BloomFlash & $\checkmark$ & $\times$ & $\checkmark$ &$\times$ & Hierarchical & RAM \& Flash & Database\\

Lu et al~\cite{Lu} & Forest-structured BF design (FBF) & $\checkmark$ & $\times$ & $\checkmark$ & $\checkmark$ & Forest Structured & RAM \& Flash & Database\\

Lee et al. \cite{Lee}  & Bloomjoin &$\checkmark$ & $\times$ & $\times$ & $\times$ & Conventional & RAM \& HDD & MapReduce Join\\


Wang et al. \cite{Wang}  & Bloom Filter Tree & $\checkmark$ & $\times$ & $\checkmark$& $\times$ & Hierarchical & RAM \& HDD & Auditing\\



BloomStore \cite{GLu}  & BloomStore & $\checkmark$ & $\times$ &$\checkmark$ & $\checkmark$ & Chain & RAM \& Flash & Key/Value deduplication \\

DBLK \cite{DBLK} & Multilayer Bloom filter & $\checkmark$ &  $\times$ & $\checkmark$ & $\times$ & Hierarchical& RAM \& HDD & Deduplication \& Compression \\

Blasco et al. \cite{Blasco} & Conventional & $\checkmark$ & $\times$ & $\checkmark$ &$\times$ & $Conventional$ & Client-Server & Proof of Ownership\\

Dong et al. \cite{Dong} & Counting Bloom Filter & $\checkmark$ & $\checkmark$ & $\checkmark$ & $\times$ & Counting & Data Center & Storage Cluster \\


DBA \cite{Wei} & Dynamic Bloom filter Array  & $\checkmark$ & $\checkmark$ & $\checkmark$ & $\checkmark$ & Dynamic & General & Large scale variable membership filter\\

Guo and Efstathopoulos \cite{Guo} & Conventional &$\checkmark$ &$\times$  &$\times$  & $\checkmark$ & Conventional & RAM \& SSD & Deduplication\\

Li et al. \cite{JLi} & Conventional&$\checkmark$ & $\times$& $\times$&$\times$ &Conventional & Cloud & Storage deduplication \\

Zhang et al. \cite{Zhang} & Conventional&$\checkmark$ &$\times$ & $\checkmark$& $\times$& Conventional& Distributed System & Storage deduplication\\

I-sieve \cite{Isieve} & Conventional &$\checkmark$ & $\times$ &$\checkmark$ &$\times$ & Conventional & Cloud & Storage deduplication\\

FFBF \cite{ASingh} & Fuzzy Folded Bloom Filter &$\checkmark$ &$\times$ & $\checkmark$& $\times$ & Conventional& Cloud Computing & Bloom Filter-as-a-Service for Big Data.  \\

Khare et al. \cite{Khare} &Standard Bloom Filter &$\checkmark$ & $\times$ &$\checkmark$ &$\times$ & Conventional& Distributed System & Frequency count of sub-graph and pruning.\\

\\ \hline

\label{tab1}
\end{longtable}
\end{landscape}
\twocolumn
\subsubsection{Multidimensional Bloom Filter}
Unlike conventional Bloom Filter, Multidimensional Bloom Filter (MDBF) uses a multidimensional Bloom Filter array. The array may be 2D, 3D, or many dimensions to reduces the false positive probability. Less work has been done in this MDBF.
\subsubsection{Compressed Bloom Filter}
Compressed Bloom Filter \cite{Mitz} is very space efficient Bloom Filter. However, there is a trade-off between space consumption and false positive on compressed Bloom Filter. The trade-off is- ``more compressing results high false positive probability''.
\section{Data Storage}
\label{DS}
Big Data consists of a variety of data, namely, structured, semi-structured and unstructured data. About 90\% of the data are unstructured in Big Data. Moreover, the data are humongous in size. It is very tough task to store, process and manage. The data size is increasing daily basis due to involvement of people in new technology. Smart Phone, for instance. People are generating data through various ways, namely, IoT, Modern Healthcare, Web, Social Media, and business. Thus, deduplication plays a vital role in preventing duplicate data in Big Data paradigm. Bloom Filter is a tiny data structure to manage the mammoth size of data to filter out the duplicate data.

Table ~\ref{tab1} extrapolates and compares the various features of the recent deployment of Bloom Filter in Big Data storage system. In the comparative study, we have found that Bloom Filter has been deployed for various purposes. 

\subsection{Flash memory}
Interestingly, tape drives are used to backup a bulk size of data. Tape drives are cheap and high storage capacity. However, tape drives are very slow. Therefore, non-working set of data are dumped in data warehouse using the tape drive. A working set of data is stored in HDD nowadays. The active working set of data is stored in SSD/PICe-PCM/NVMe and passive working set of data are stored in HDD. Processing speed of RAM is 1000 times faster than HDD and 100 times faster than SSD. Cost of RAM is higher than SSD and HDD. SSD is costlier than HDD. Thus, SSD/PCIe-PCM/NVMe replace the HDD in Big Data era. Most of the datacenters are upgrading their storage to SSD/Flash memory. However, HDD is also used for various purposes.

BloomFlash \cite{BloomFlash} is a design of a NAND Flash memory based storage system for very large scale data size. BloomFlash consider the case of very large Bloom Filter that does not fit in RAM. BloomFlash is cleverly designed to maximize the random writes and random read throughput. In-place bit update is costlier in SSD than RAM. Hence, BloomFlash employs a lazy update policy of Bloom Filter in Flash memory. In addition, BloomFlash augment tiny Bloom Filter in every flash page to localize the reads and writes on Flash memory. Thus, BloomFlash is able to scale massive amount of data. 

Moreover, Lu et al. \cite{Lu} devises a forest structured Bloom Filter (FBF) for Flash memory storage. FBF deals with RAM and Flash memory. Similar to BloomFlash \cite{BloomFlash}, Bloom Filter is stored in RAM and Flash memory. Write and read operations are instantly done in RAM-based Bloom Filter. On the contrary, write operation request into Flash memory is delayed and it is buffered in memory to update later on. The update to Bloom Filter is performed periodically to maximize the write performance. Random write operation takes more time than random read operation. Because, write operation requires erasing the data present in the Flash memory. Hence, it is a clever way to wait for some time to write information into the Flash memory. However, read operations are instantly accessed in the Flash memory. The FBF is forest-structured Bloom Filter that initially resides in RAM and expands the Bloom Filter to Flash memory. The Bloom Filter sizes are small and fixed, and thus, it is very easy to grow. The lookup time complexity of FBF is $O(\log_bm)$ due to forest-structured, where $b$ is the total branches (degree of the tree) and $m$ is the overall FBF size. The FBF is very useful in very large scale deduplication of data.

Similarly, BloomStore\cite{GLu} is a key/value storage system using NAND Flash memory. BloomStore is similar to FBF \cite{Lu}. Each physical node spawns several BloomStore within its key ranges. The key ranges are disjoint to each other. The BloomStore addresses an efficient key/value storage in Flash memory. Conventional hash consumes more storage space as compared to Bloom Filter variant. Thus, BloomStore implements the key/value Flash memory storage with a chain architecture of Bloom Filters.

\subsection{Cloud storage}
BigTable \cite{BigTable} deploys Bloom Filter to avoid unnecessary disk accesses. The disk access takes huge times. Therefore, Bloom Filter is checked for membership for a query before being accessed to the disk. If Bloom Filter denies, then no hard disk access is performed. In addition, keys are stored in a Sorted String Table (SST). SST stores data in RAM and HDD. However, the access request to SST also a time consuming process. If a data does not exist, then lookup time in SST is wasted. Thus, Bloom Filter enhances drastically the Big Data performance. Similarly, Cassandra also deploys Bloom Filter to avoid unnecessary disk accesses \cite{Lakshman}. Moreover, Lahiri et al. \cite{Lahiri} deploys Bloom Filter to filter the duplicate data in querying oracle databases. 

Kaur and Sood \cite{Kaur} proposes a resource management system in Big Data streaming. Bloom Filter is used to determine the variety of data. The author considers four types of streaming data, namely, text data, audio data, video data, and image data. Four Bloom Filters are deployed to separate the types of data in the Big Data streaming which greatly boosted up the streaming performance, and reduces the memory consumption.

Gessert et al. \cite{Gessert} implements Orestes, Database-as-a-Service, by deploying Bloom Filter to achieve low latency. Bloom Filter is extremely useful in the web browser cache filtering. Hence, Orestes \cite{Gessert} uses Bloom Filter web browser cache filtering to prevent from reading the stale data. Gessert et al. \cite{Gessert} devises a new variant of Bloom Filter, called Bloom-Filter-Bound (BFB) to provide cache consistency to Orestes. The key focus of BFB is to prevent the reading stale data from cache. BFB guarantees to provide recent object versions. BFB enhances the performance of Orestes and ensure cache consistency.  

Moreover, Li et al. \cite{JLi} constructs secure Cloud storage where Bloom Filter is used for trapdoor indexing. The trapdoors are encrypted using secret keys of a user. Keywords are encrypted and then inserted into the indexing. A user computes hash value of a file to be uploaded, and sends it to the Cloud storage server. The user must proof the ownership of the file by answering the challenge given by the server. The Bloom Filter is examined for the existence of the user request. 

In addition, Droplet \cite{Zhang} is distributed storage system that employs Bloom Filter for fingerprint indexing. Droplet avoids unnecessary disk lookup by Bloom Filter which boost up the performance. Zhang et al. \cite{Zhang} claims that Droplet is able to avoid 90\% of unnecessary disk lookup by deploying Bloom Filter. 

Moreover, Bloom Filter is also deployed in Bioinformatics, for instance, Genome project. Melsted and Pritchard \cite{Melsted} deploys Bloom Filter for k-mer DNA sequencing. A single DNA data are very large to store, process and manage. Sequencing DNA is challenging task. $de~Bruijn$ graph represents k-mer as a node and billions of such k-mer nodes are interconnected to form the $de~Bruijn$ graph. Bloom Filter is deployed to discover the existence of k-mer in $de~Bruijn$ graph. All k-mers are inserted into Bloom Filter which is greatly compacted for storing information. 

Similarly, Chikhi and Rizk \cite{Chikhi} also construct $de~Bruijn$ graph through deploying Bloom Filter. ABySS 2.0 \cite{Abyss} uses the Bloom Filter for resource efficient assemble of large genomes. Use of Bloom Filter reduces the overall memory requirements and enables to assemble large genomes on a single machine. It is used to represent the de Brujin Graph. Likewise, BLESS \cite{Heo} is also DNA sequencing using Bloom Filter. However, BLESS re-correct the error caused by false positive for sequencing. However, the false positive has been an open problem for Bloom Filter since its inception.

Moreover, demands of Bloom Filters is increasing for Big Data research. Therefore, Singh et al. \cite{ASingh} designs a Bloom Filter for hosting in Cloud Computing environment, and enables Bloom Filter-as-a-Service. Singh et al. \cite{ASingh} devices a novel fuzzy Bloom Filter, called Fuzzy-folded Bloom Filter (FFBF). FFBF is deployed to design Bloom Filter-as-a-Service. It incorporates a very large amount of the data which enables ease of computing for the Big Data using Bloom Filter. 

\section{Future direction}
\label{FD}
As we have experienced, SSD/Flash memory is replacing HDD, and thus, HDD is used for backup purposes. The HDD will be completely replaced by SSD/Flash memory in near future. Therefore, the direction of Bloom Filter is designing an in-memory Bloom Filter and Flash-based Bloom Filter. The Flash memory is faster than HDD, however, slower than RAM. But, the RAM space is precious, and a Bloom Filter alone cannot consume entire spaces of RAM. Thus, designing RAM and Flash-based  Bloom Filter is a challenge. In addition, Bloom Filter requires few bit update upon insertion of an element. The Flash-based Bloom Filter suffers from such update, because random write consume more time in Flash memory. Thus, it requires a clever way to design such kind of Bloom Filter. As we have discussed, the lookup time also depends on the structure of Bloom Filter, for example, forest structured and chained structured Bloom Filter. 

Moreover, Bloom Filter is an unintelligent filter, but it is time to bring in the smart Bloom Filter that can learn. Incorporating machine learning and AI can make Bloom Filter even better than what we expect. The future of Bloom Filter will act as a smart filter, that can identify patterns. On the other hand, Bloom Filter is also deployed interdisciplinary computing area. Moreover, $de~Bruijn$ graph also be constructed in Flash memory using Bloom Filter which will greatly impact the DNA sequencing performance. 

Bloom Filter is inapplicable in exact query requirements. For instance, real-time system. Moreover, Bloom Filter can also be deployed in a password management system which will greatly compact the storage space. However, false positive is the key barrier of the password management system. Moreover, Bloom Filter is also deployed in IoT, Fog Computing, and recent emerging technology to reduce the space consumption. On the other hand, scalability of Bloom Filter is also a big challenge. However, it is addressed using Flash memory. It is observed that there is a challenge in designing scalable in-memory Bloom Filter that is more compact to accommodate a larger number input. Besides, forest structured and chain structured Bloom Filter take high lookup cost. Distributed and parallel approach of Bloom Filter designing is also called for to reduce lookup cost.

\section{Conclusion}
\label{Con}
In this article, we feature the applicability of Bloom Filter in Big Data storage system. We have found that Flash memory is most promising research area for integrating Bloom Filter to build Big Data storage system. We also demonstrated that scalability of Bloom Filter is a serious issue which can be achieved by implementing Bloom Filter in Flash memory. The random write of Flash memory is faster than writing into disk. However, writing a bit to Flash memory requires erasing the entire pages and write it again. Hence, random write is slower in Flash memory than RAM. On the contrary, the RAM size is limited and deduplication requires huge memory space. Therefore, implementing Bloom Filter in Flash memory. Nevertheless, there is a tradeoff between performance and space. 

In this review, we find that most of the Big Data storage system deploy Bloom Filter to avoid unnecessary disk access which boost up the system performance dramatically. Moreover, Bloom Filter also enhances the performance of DNA sequencing. The billions of nodes in $de~Bruijn$ graph is really arduous task to process without Bloom Filter. Finally, we conclude that a tiny data structure takes a lion's share in performance improvement in Big Data.

\bibliographystyle{IEEEtran}
\bibliography{mybibfile}

\end{document}